\documentclass[epj]{svjour}
\usepackage{graphicx}       
\usepackage{bm}         
\usepackage{amsfonts}
\usepackage{amsmath}        

\begin{document}

\title{Entanglement preparation using symmetric multiports}
\author{Thomas Brougham\inst{1}\thanks{Corresponding author email:thomas.brougham@gmail.com}, Vojt\v{e}ch Ko\v{s}t'\'{a}k\inst{1}, Igor Jex\inst{1}, Erika Andersson\inst{2} and Tam\'{a}s Kiss\inst{3}}
\institute{Department of Physics, FNSPE, Czech Technical University in Prague, B\v{r}ehov\'{a} 7, 115 19 Praha 1 - Star\'{e} M\v{e}sto, Czech Republic \and
SUPA, Department of Physics, School of EPS, Heriot-Watt University, Edinburgh \and
Department of Quantum Optics and Quantum Information, Research Institute for Solid State Physics and Optics, Hungarian Academy of Sciences,
Konkoly-Thege Mikl\'{o}s \'{u}t 29-33, H-1121 Budapest, Hungary}

\abstract{We investigate the entanglement produced by a multi-path interferometer that is composed of two symmetric multiports, with phase shifts applied to the output of the first multiport.  Particular attention is paid to the case when we have a single photon entering the interferometer.  For this situation we derive a simple condition that characterize the types of entanglement that one can generate.  We then show how one can use the results from the single photon case to determine what kinds of multi-photon entangled states one can prepare using the interferometer.
}
\PACS{{42.50.p}{Quantum Optics}\and
{03.67.Bg}{Entanglement Production}}
\titlerunning{Entaglement preparation using symmetric multiports}
\authorrunning{T. Brougham, V. Kostak et al.}

\maketitle

\section{Introduction}  
\label{secI}
Entanglement is perhaps the most important and enigmatic feature of quantum mechanics.  It is at the heart of quantum computing \cite{qcomp1,qcomp2,nch} and has many other applications within the field of quantum information \cite{teleport,densecode,qkd}.  The task of generating entangled states is thus of practical as well as fundamental importance.  A simple way of generating entanglement, within optics, is to use a balanced beam splitter \cite{kim}.  A straightforward, but powerful, generalization of this would be to use a Mach-Zehnder interferometer \cite{loudon}.  Using this simple setup one can generate many different  entangled two mode states.  For instance, this setup allows one to generate all possible two mode, single photon entangled states, i.e all states of the form $a|1\rangle|0\rangle+b|0\rangle|1\rangle$\footnote{It is sometimes claimed that such states are not truly entangled.  The states can, however, be used to violate a suitable Bell inequality \cite{bell,bell2} and thus are entangled in the conventional sense.  Furthermore, the question: `what is the photon entangled with?', can be addressed by noting that it is the two modes of the electromagnetic field that are entangled, not the photon.  
}.  

It would seem natural to try to generate multipartite entangled states using a similar linear optical setup.  
This can be achieved using two symmetric multiports \cite{bm,zeilinger,zeilinger2,WRWZ,walker,stig,lb,mp} and phase shifts.  Interferometers such as this have been studied previously \cite{genmp,mp2,mp} and have been realized experimentally \cite{WRWZ}.  The advantage of using this setup over more general schemes, such as \cite{genmp}, is that it can be realized straightforwardly using commercially available fiber couplers \cite{WRWZ}.  The previous work on multi-path interferometers have focused on showing that a specific state or class of states can be generated using the setup.  There has, however, not been a systematic study of the states that these interferometers can produce.  In this paper we will give a general treatment that determines all possible states that can be generated for single photon input.  These results will then be used to show how one can classify the possible output states for two photons entering the apparatus.  

The results will allow us to answer a number of basic questions regarding multi-path interferometers.  For example, the Mach-Zehnder interferometer allows one to generate all possible single photon (two mode) entangled states, does this hold in the higher dimensional case?  We will demonstrate that this is not the case.  Similarly, if one has a multi-path interferometer, can the setup function as a lower dimensional multi-path interferometer?  For example, by ignoring one of the paths, can we choose the phase shifts so that a three path interferometer functions as a two path interferometer?  
Finally, we will investigate what kinds of two photon entangled states one can generate.

\begin{figure}
\center{\includegraphics[width=8cm,height=!]
{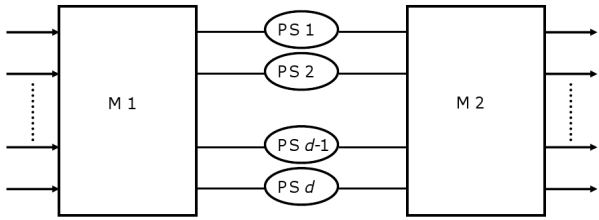}}
\caption{A schematic diagram of a multi-path interferometer.  The elements M1 and M2 are symmetric multiports, while  PS1, PS2,..., PS $d-1$ and PS $d$ are phase shifts.}
\label{fig1}
\end{figure}

\section{Multiports and multi-path interferometers}
\label{secII}
Beam splitters are ubiquitous within quantum optics and are an essential component of many quantum information protocols \cite{nch,qip,qip2,perina}.  Because of this it is interesting to consider generalizations of balanced beam splitters to the case where we have $d$ inputs and $d$ outputs.  Such devices are known as symmetric multiports.  If a single photon enters one of the inputs of a symmetric multiport, then we have probability $1/d$ of finding it at a given output.  Without loss of generality, we can describe the  action of a symmetric multiport on a single photon input by using the discrete Fourier transform.  This means that the action of the beam splitter on the input state  $|\psi\rangle=\sum_k{b_k|1_k\rangle}$, where $|1_k\rangle=|0\rangle^{\otimes k}|1\rangle|0\rangle^{\otimes d-1-k}$, will be $|\psi\rangle\rightarrow|\psi'\rangle=\sum_{j=0}^{d-1}{c_j|1_j\rangle}$, where
\begin{eqnarray}
c_j=\sum_k{\frac{1}{\sqrt{d}}\exp\left(\frac{2\pi i jk}{d}\right)b_k}.
\end{eqnarray}

Using symmetric multiports we can describe a generalization of the Mach-Zehnder interferometer, which has $d$ inputs and $d$ outputs.  The interferometer will consist of two symmetric multiports, with phase shifts performed on the outputs of the first multiport.  This can be represented mathematically by the following unitary operator
\begin{equation}
\label{unitary}
\hat U=\hat F\hat P\hat F,
\end{equation}
where $\hat F$ is the discrete Fourier Transform, which we define here as $\hat F=\sum_{m,n}{\exp(2\pi imn/d)/\sqrt{d}|1_m\rangle\langle 1_n|}$, and $\hat P$ represents the phase shifts acting on each mode, hence $\hat P=\sum_k{\lambda_k|1_k\rangle\langle 1_k|}$, with $|\lambda_k|=1$.  Suppose that a single photon enters one of the inputs of the interferometer.  Without loss of generality we shall assume that the photon enters the first input, hence the input state is $|1_0\rangle$.  The output state will thus be $\hat U|1_0\rangle=|\psi\rangle=\sum_m{c_m|1_m\rangle}$, where
\begin{equation}
\label{c}
c_m=\frac{1}{d}\sum_n{\lambda_n\exp\left(\frac{2\pi imn}{d}\right)}.
\end{equation}
Equation (\ref{c}) can be expressed in the compact form ${\bf c}=F{\boldsymbol\lambda}/\sqrt{d}$, where the vectors ${\bf c}$ and $\boldsymbol{\lambda}$ each have components $c_k$ and $\lambda_k$ respectively, and $F$ has elements $F_{mn}=\exp(2\pi imn/d)/\sqrt{d}$.  The results are not substantially changed by having the photon enter the interferometer via a different input port.  In particular, the set of possible output states that one can generate is not dependant on which input port the photon enters.  For example, if we generate the state $|\psi\rangle$ with the photon entering via the $m$-th input and with phase shifts ${\boldsymbol \lambda}$, then we can also generate this state by having the photon enter the first input and using phase shifts that are a permutation of ${\boldsymbol\lambda}$.  

The question we now ask is: can a given state of the form $|\psi\rangle=\sum_k{c_k|1_k\rangle}$ be generated by the unitary transform (\ref{unitary}), when a single photon enters the first input?  If the state can be made, then the phase shifts needed to produce the state are  given by
\begin{equation}
\label{pshift}
\boldsymbol{\lambda}=\sqrt{d}F^{\dagger}{\bf c}.
\end{equation}
The fact that $|\lambda_k|=1$, for $k=0,1,...,d-1$, implies that the state can be produced if and only if $|(F^{\dagger}{\bf c})_k|=1/\sqrt{d}$.  It will be convenient to write this condition in a different form.  By using equation (\ref{pshift}) together with the fact that $|\lambda_k|^2=1$, we can obtain
\begin{equation}
\label{exact}
\sum_{m, n(m\ne n)}{c_mc^*_n\exp\left(\frac{2\pi ik[n-m]}{d}\right)}=0\text{   for  }k=0, 1, 2, ...d-1,
\end{equation}  
where we have used the fact that $\sum_m{|c_m|^2}=1$.  Multiplying the left hand side of (\ref{exact}) by $\exp(2\pi ipk/d)$ and then summing over $k$ leads to the expression
\begin{equation}
\sum_{k,m,n(m\ne n)}{c_mc^*_n\exp\left(\frac{2\pi ik[n-m+p]}{d}\right)}=0.
\end{equation}
The only cases when the summation $\sum_k{\exp(2\pi ik[n-m+p]/d)}$ will not be zero is when $n-m+p=0$ or $n-m+p=d$.  This observation allows us to obtain the following new set of equations
\begin{equation}
\label{exact2}
\sum_{m=0}^{d-1}{c_mc^*_{m-p}=0};\;\text{for }p=1,2,...,d-1,
\end{equation}
where the indices of each $c$ term are modulo $d$, hence $c_{x+d}=c_x$. It is clear that a solution of (\ref{exact}) will also be a solution of (\ref{exact2}).  With some thought one can easily establish that the converse is true and thus equation (\ref{exact2}) is equivalent to equation (\ref{exact}).


\section{Characterizing the entanglement generated from a $d$-port interferometer}  
\label{secIII} 

While equation (\ref{exact2}) totally characterizes the possible output states, there are situations where it is desirable to have an alternative condition.  This can be illustrated by the following example.  Suppose one wanted to generate an entangled state with coefficients given by ${\bf c}$, which is not a solution of (\ref{exact2}).  Let us further suppose that it is possible to generate a state with coefficients ${\bf c}'$, where $|c'_j|=|c_j|$, $j=0,1,...,d-1$.  It is clear that we can transform this particular state to the desired one by simply making appropriate phase shifts to the output of the interferometer.  One could thus say that these two entangled states are equivalent.  The sense in which the states are equivalent can be made precise by considering the set of concurrences, $\{C_{m,n}\}$, between modes $m$ and $n$ \cite{wootters}\footnote{In general, multipartite entanglement cannot be fully characterized using only the concurrences between each particle \cite{counterex,counterex2}.  The entanglement of a single photon state, however, can be fully characterized using the set of concurrences \cite{jaroslav}.}.  It has been shown that for the single photon case, the concurrence between two modes $m$ and $n$ reduces to \cite{jaroslav}
\begin{equation}
\label{concur}
C_{m,n}=2|c_m||c_n|. 
\end{equation}
The probability distribution for finding the photon at a particular output port, $P_m=|c_m|^2$, is thus enough to characterise the entanglement between any two modes of the output state.  This raises the question of whether we can design the phase shifts of the interferometer so that we can produce a state with a specific probability distribution $P_m$.  For $d=2$, it is clear that we can create states with any desired output probability distribution, i.e. we can prepare states of the form $a|10\rangle+b|01\rangle$, where $|a|^2$ can have any value between 0 and 1.  For multi-path interferometers with $d$ outputs, we will find that there are limits to the possible output probability distributions.  

In order to find general conditions on the allowed output probability distributions, we will use a simple geometrical representation of equation (\ref{exact2}).  One can obtain this geometrical picture of (\ref{exact2}) by plotting the terms $c_mc^*_{m-p}$ on the complex plane.  Each term $c_mc^*_{m-p}$ will correspond to directed lines of length $|c_m||c_{m-p}|$.  The condition that the sum of these terms must equal zero, will correspond to the directed lines forming a closed polygon in the complex plane.  It is clear that for any polygon the length of one side must be less than the sum of the lengths of the other sides.  This fact together with equation (\ref{exact2}) lead to the following condition 
\begin{equation}
\label{gennec}
|c_m||c_{m-p}|\le\sum_{k(k\ne m)}{|c_k||c_{k-p}|},\;m,p=1,2,...,d-1.
\end{equation}
One can compare a specific photon probability distribution against the necessary condition (\ref{gennec}).  If the probability distribution violates the set of inequalities, then we cannot prepare a state where the probability of finding the photon in a particular mode is the same as the specified probability distribution.  
The concurrence between any two modes of the output state is given by (\ref{concur}); we can use this result to re-cast equation (\ref{gennec}) in the following form 
\begin{equation}
\label{necconc}
C_{m,m-p}\le\sum_{k(k\ne m)}{C_{k,k-p}},\;m,p=1,2,...,d-1,
\end{equation}
where the indices of the concurrence are taken to be modulo $d$, i.e. $C_{m+d,n+d}=C_{m,n}$.  The inequalities (\ref{necconc}) place bounds on how entangled  two modes of the output state can be.  

While we have shown that both equations (\ref{gennec}) and (\ref{necconc}) are necessary conditions, we have not addressed the question of whether they are also sufficient.  It will be proved in section \ref{secIV} that for $d=3$, the conditions are also sufficient.  For $d>3$ the question is still open, however, numerical investigation suggest that equations (\ref{gennec}) and (\ref{necconc}) are not sufficient for $d=4$ and 5.

Some consequences of equation (\ref{necconc}) will be examined.  Suppose we want to construct a state where only two modes were entangled.  This would correspond to the concurrence being zero between all but the two modes that are entangled.  Let us denote the two entangled modes as $a$ and $b$.  We thus have that $C_{a,b}=C_{b,a}>0$, while $C_{m,n}=0$ for all other choices of $m$ and $n$.  When the left hand side of equation (\ref{necconc}) is not equal to $C_{a,b}$ or $C_{b,a}$, then the inequality (\ref{necconc}) is trivially satisfied.  Consider the situation where the left hand side of (\ref{necconc}) equals $C_{a,b}$.  Clearly the integer $p$ must equal $|a-b|$.  If the two modes $a$ and $b$ are to be entangled then $C_{a,b}>0$, which implies that $\sum_k{C_{k,k+b-a}}>0$, where $k\ne a$.  This condition can only be fulfilled when the summation $\sum_k{C_{k,k+b-a}}$, $k\ne a$, contains the term $C_{b,a}$, hence $C_{k,k+b-a}=C_{b,a}$, which implies that $[2b-a]\mod d=[a]\mod d$.  If this last expression is true then we see that $|(2b-a)-a|=2|b-a|=nd$, where $n$ is a positive integer.  The fact that $a$ and $b$ belong to the set of integers $\{0,1,2,...,d-1\}$ means that $|b-a|<d$ and thus $2|b-a|=d$, which implies that $d$ must be even.  If $d$ is not even then $C_{a,b}\le 0$ and the two modes will not be entangled.  We thus have the following result. An interferometer of the type described in equation (\ref{unitary}) cannot create a state where only two modes are entangled, when the interferometer has an odd number of input and output ports.

For $d$ even, it is possible to prepare states where only two modes are entangled.  For example the state $|\psi\rangle=i|c_a||1_a\rangle+|c_{b}|1_b\rangle$, where $b=a+d/2$ and $|c_a|^2+|c_{b}|^2=1$, can be produced.  This can be easily verified using equation (\ref{exact2}).  

\section{Entanglement generation using a 3-port interferometer}
\label{secIV}
The set of inequalities (\ref{gennec}) are necessary conditions upon us generating a state with a photon probability distribution $|c_m|^2$.  When our interferometer has three inputs and outputs, it can be shown that these conditions are also sufficient.  
We will now give a proof of this fact.  The approach that we shall use is constructive in that it leads to a method for calculating the phase shifts need to engineer the desired state.  

When our interferometer has 3 input and output ports, equation (\ref{exact2}) is a set of two equations, each of which is the complex conjugate of the other.  Equation (\ref{exact2}) thus reduces to a single equation 
\begin{equation}
\label{3d}
c_0c_2^*+c_1c_0^*+c_2c_1^*=0.
\end{equation}
The same argument that leads to equation (\ref{gennec}) will yield the following set of inequalities
\begin{eqnarray}
\label{triangle}
|c_0||c_2|\le|c_0||c_1|+|c_1||c_2|,\nonumber\\
|c_1||c_0|\le|c_0||c_2|+|c_1||c_2|,\nonumber\\
|c_1||c_2|\le|c_0||c_2|+|c_0||c_1|.
\end{eqnarray}
If the inequalities (\ref{triangle}) are not satisfied, then we cannot prepare a state with coefficients $\{c_m\}$.  
Suppose that we have a state with coefficients $\{c_m\}$, which satisfy the inequalities (\ref{triangle}), but not equation (\ref{3d}).  In this case plotting the complex numbers $c_0c_2^*$, $c_1c_0^*$ and $c_2c_1^*$, on the complex plane will not give a closed shape.  The fact that the lengths of the three directed lines satisfy the triangle inequalities, equation (\ref{triangle}), means that they could be rotated so that they  form a triangle.  This is equivalent to changing the phases of the complex numbers $c_m$.  We will now show that when $d=3$, we can always choose the phases so that a set $\{|c_m|\}$ that satisfies (\ref{triangle}) will be solutions of equation (\ref{3d}).

\begin{figure}
\center{\includegraphics[width=8cm,height=!]
{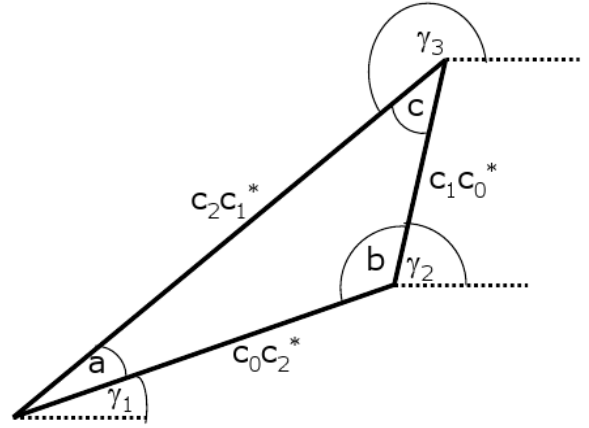}}
\caption{A diagram showing the terms $c_0c_2^*$, $c_1c_0^*$ and $c_2c_1^*$ represented as lines in the complex plane, where the lines form a triangle.}
\label{fig2}
\end{figure}

Let $\phi_m$ be the phase of $c_m$, i.e. $c_m=|c_m|e^{i\phi_m}$ and let  $\gamma_1=\phi_0-\phi_2$, $\gamma_2=\phi_1-\phi_0$ and $\gamma_3=\phi_2-\phi_1$.  The condition that a set of coefficients $\{c_m\}$ satisfy equation (\ref{3d}) can be represented geometrically as saying that they must form a triangle in the complex plane, as shown in figure \ref{fig2}.  
The values of the angles $a$, $b$ and $c$ (that are defined in figure 2) can be found using the cosine rule
\begin{eqnarray}
\label{cosrule}
\cos(a)=\frac{|c_0|^2|c_2|^2+|c_1|^2|c_2|^2-|c_1|^2|c_0|^2}{2|c_0||c_1||c_2|^2},\nonumber\\
\cos(b)=\frac{|c_0|^2|c_2|^2+|c_0|^2|c_1|^2-|c_1|^2|c_2|^2}{2|c_0|^2|c_1||c_2|},\nonumber\\
\cos(c)=\frac{|c_0|^2|c_1|^2+|c_1|^2|c_2|^2-|c_0|^2|c_2|^2}{2|c_0||c_1|^2|c_2|}.
\end{eqnarray}
Using some simple geometry we find that
\begin{eqnarray}
\label{app2}
a=\gamma_1-\gamma_3+\pi,\nonumber\\
b=\gamma_2-\gamma_1-\pi,\nonumber\\
c=\gamma_3-\gamma_2+\pi. 
\end{eqnarray}
For the sake of simplicity we shall take $\phi_0=0$, this is equivalent to making a global phase change on the output state $|\psi\rangle=\sum_m{c_m|1_m\rangle}$.  It is now straightforward to see that we can take the other phases to be
\begin{eqnarray}
\label{phasesb}
\phi_1=\frac{1}{3}\left(2b+a+\pi\right),\nonumber\\
\phi_2=\frac{1}{3}\left(b-a+2\pi\right).
\end{eqnarray}
This is equivalent to preparing the state
\begin{equation}
\label{stateb}
|\psi\rangle=|c_0||1_0\rangle+|c_1|xe^{i(a+2b)/3}|1_1\rangle+|c_2|x^2e^{i(b-a)/3}|1_2\rangle,
\end{equation}
where $x=\exp(\pi i/3)$.  

The problem of determining whether it is possible to engineer a state with a particular probability for finding the photon in a given mode, has been solved for the case of $d=3$.  The inequalities (\ref{triangle}) represent necessary and sufficient conditions for one to engineer a state with the given probability distribution.  



Equation (\ref{triangle}) tells us whether we can engineer a state with a particular property; this still leaves open the problem of how to actually prepare the state, i.e. what phase shifts are required.  This problem can be resolve by using equation (\ref{stateb}) together with equation (\ref{pshift}) to obtain the following phase shifts
\begin{eqnarray}
\label{phasechoice}
\lambda_0&=&|c_0|+|c_1|xe^{i(a+2b)/3}+|c_2|x^2e^{i(b-a)/3},\nonumber\\
\lambda_1&=&|c_0|+|c_1|x^*e^{i(a+2b)/3}+|c_2|(x^2)^*e^{i(b-a)/3},\nonumber\\
\lambda_2&=&|c_0|-|c_1|e^{i(a+2b)/3}+|c_2|e^{i(b-a)/3},
\end{eqnarray}
where $a$ and $b$ are determined from equation (\ref{cosrule}).  It is clear that this choice of phase shifts is not unique and that different phase shifts could be used to generate the same output photon probability distribution.

\section{Multi-photon entanglement generation}
\label{secV}
From the perspective of quantum information, multi-photon entangled states are of great interest.  For this reason we will investigate the types of multi-photon entanglement states that one can generate using the setup described in figure 1.  

Let $\hat a_m$ and $\hat b_m$ be bosonic annihilation operators that acts on the $m$-th mode, i.e. $\hat a_m|0\rangle=|1_m\rangle$, $\hat b_m|0\rangle=|1_m\rangle$, $[\hat a_i,\hat a^{\dagger}_j]=\delta_{ij}\hat 1$ and $[\hat b_m,\hat b^{\dagger}_n]=\delta_{mn}\hat 1$.  The operators $\{\hat a_m\}$ will be associated with the input modes of the field, while $\{\hat b_m\}$ are associated with the output modes of the field.  The fact that symmetric multiports are linear optical devices means that output operators $\{\hat b_m\}$ are related to the output operators $\{\hat a_m\}$ by the linear equations
\begin{equation}
\label{linab}
\hat b_m=\sum_k{V_{mk}\hat a_k},
\end{equation}
where $V_{mk}$ are the elements of a unitary matrix, i.e. $\sum_k{V_{mk}}$ $V^*_{nk}=\delta_{mn}$.  In order for us to make a connection between $V$ and equation (\ref{unitary}), we will consider the situation where a single photon enters the interferometer.  The input state will be $\hat a^{\dagger}_m|0\rangle$, while the output state is $\sum_k{V_{mk}\hat b^{\dagger}_k|0\rangle}$, where we have used equation (\ref{linab}).  When we have a single photon input, we can use equation (\ref{unitary}) to express the output state as $\hat U|1_m\rangle=\sum_k{U_{km}\hat b^{\dagger}_m|0\rangle}$, hence $V_{mk}=U_{km}$ i.e. $U=V^{T}$.  Equation (\ref{c}) shows that $U_{m0}=(\sum_k{F_{mk}\lambda_k})/\sqrt{d}=c_m$.  
For $n\ne 0$, a straightforward calculation shows that $U_{mn}=c_{m+n}$, where we take the index of $c$ to be modulo $d$, i.e. $c_{x+d}=c_{x}$.  Combining this with the previous results we find that
\begin{equation}
\label{transferm}
V_{mn}=c_{n+m}.
\end{equation}
The single photon results, such as equation (\ref{gennec}), can be used to place restrictions on the types of multi-photon entanglement that we can generate.


We will illustrate the previous theory by investigating the entanglement that one can generate from a multi-path interferometer with three inputs and outputs.  The first situation that we will examine is when we have two photons that both enter through the same input.  Without loss of generality, we can assume that the photons enter through input 0, i.e. our input state is $|200\rangle=(\hat a_0^{\dagger})^2|0\rangle/\sqrt{2}$.  Using equation (\ref{transferm}) together with some simple algebra we find the following general form for the output states
\begin{eqnarray}
\label{2photon2}
|\psi\rangle=N\bigl(c^2_0|200\rangle+c^2_1|020\rangle+c^2_2|002\rangle\nonumber\\
+2c_0c_1|110\rangle+2c_0c_2|101\rangle+2c_1c_2|011\rangle\bigr),
\end{eqnarray}
where $N$ is a normalization constant.  Suppose we want to determine whether or not we can generate a two photon entangled state that has a particular probability distribution for finding the photons in the outputs.  For example, the state $|\psi'\rangle=\alpha_0|200\rangle+\alpha_1|020\rangle+\alpha_2|002\rangle+\alpha_3|110\rangle+\alpha_4|101\rangle+\alpha_5|011\rangle$, has the probability distribution 
\begin{eqnarray}
\label{2photonprob}
P_{200}=|\alpha_0|^2,\;P_{020}=|\alpha_1|^2,\;P_{002}=|\alpha_2|^2,\nonumber\\
P_{110}=|\alpha_3|^2,\;P_{101}=|\alpha_4|^2,\;P_{011}=|\alpha_5|^2.
\end{eqnarray}
If a state with this probability distribution is to be generated by having both photons enter a single input, then from equation (\ref{2photon2}) it is clear that $|c_k|=\sqrt{|\alpha_k|/N}$, for $k=0,1,2$, which implies that 
\begin{eqnarray} 
\label{2photon2nec}
|\alpha_3|=2\sqrt{|\alpha_0||\alpha_1|},\; |\alpha_4|=2\sqrt{|\alpha_0||\alpha_2|}\;\&\; |\alpha_5|=2\sqrt{|\alpha_1||\alpha_2|}.\nonumber\\
\end{eqnarray}
If these three necessary conditions are not satisfied, then the entangled state cannot be generated by having two photons enter the same input of a three port symmetric interferometer.  When the conditions (\ref{2photon2nec}) are satisfied, then we must check that the values of $|c_k|$, defined by $|c_k|=\sqrt{|\alpha_k|/N}$, satisfy the set of inequalities (\ref{triangle}).  If this last condition is met, then we will be able to generate an entangled state with the given probability distribution.  The phase shifts need to generate the state will be the same as the phase shifts that generate the single photon state $\sum_k{c_k|1_k\rangle}$, which are given in equation (\ref{phasechoice}). 

The utility of the previous results are best illustrated by two quick examples.  The output photon probability $P_{200}=P_{020}=P_{002}=1/9$ and $P_{110}=P_{101}=P_{011}=2/9$, will violate the conditions (\ref{2photon2nec}) and thus we cannot generate a  state that have this output probability distribution.  By contrast, the output probability distribution $P_{200}=P_{020}=P_{002}=1/15$ and $P_{110}=P_{101}=P_{011}=4/15$ will satisfy both equation (\ref{2photon2nec}) and the inequalities (\ref{triangle}). Using equations (\ref{cosrule}) and (\ref{phasechoice}) together with some simple algebra, we find that the phase shifts needed to obtain an entangled two photon state with the desired probability distribution are $\lambda_0=e^{i\pi/2}$ and $\lambda_1=\lambda_2=e^{-i\pi/6}$. 

We can also consider generating entangled state by having photons enter through two different inputs.  Without loss of generality, we assume that the photons enter through inputs 0 and 1, i.e. the input state is $|110\rangle=\hat a^{\dagger}_0\hat a^{\dagger}_1|0\rangle$.  From equation (\ref{transferm}) we find that the general form for the output state is
\begin{eqnarray}
\label{2photon1}
|\psi\rangle=N\Bigl[c_0c_1|200\rangle+c1c_2|020\rangle+c2c0|002\rangle\nonumber\\
+(c_0c_2+c1^2)|110\rangle+(c_1c_2+c_0^2)|101\rangle+(c_0c_1+c_2^2)|011\rangle\Bigr]\nonumber,\\
\end{eqnarray}
where $N$ is a normalization constant.  We can again use our previous results to determine whether a given entangled state can be generated from this setup.  Suppose that we again wish to generate an entangled state that has the probability distribution (\ref{2photonprob}).  From equation (\ref{2photon1}) it is clear that $|\alpha_0|=N|c_0||c_1|$, $|\alpha_1|=N|c_1||c_2|$ and $|\alpha_2|=N|c_0||c_2|$.  The triangle inequalities (\ref{triangle}) can again be applied to determine whether a given state is allowed.  
The problem of determining what kind of three photon entangled states one can generate using the multiport, can in principle be tackled in the same manner.  


\section{Conclusions} 
We have studied the states that one can generate using a multi-path interferometer that is composed of two symmetric multiports with phase shifts applied to the outputs of the first multiport.  For a fixed input one can obtain different output states by changing the phase shifts.  We gave a thorough treatment of the case when a single photon enters the apparatus before investigating the case of two input photons.  

For interferometers with three modes we where able to fully determine all the single photon states that one could generate in terms of a simple inequality on the concurrencies between each mode.  Furthermore, we determined the phase shifts required to generate a three mode entangled state with a prescribed output photon probability distribution.  This is significant as it means the results can be applied directly to experimental realization of three path interferometers such as \cite{WRWZ}.

The results for the single photon case, while interesting in their own right, also have relevance in determining what types of multi-photon entanglement that one can prepare.  
We were able to use the single photon results to determine what types of two photon states could be prepared.

Our results allowed us to address an number of basic questions relating to multi-path interferometers.  The first of these was whether by adjusting the phase shifts one can prepare a single photon entangled state $|\psi\rangle=\sum_m{c_m|1_m\rangle}$, such that $|c_m|^2$ realizes all possible $d$ outcome probability distributions.  For $d=2$ (i.e. the Mach-Zehnder interferometer) this is indeed the case, however, we have shown that this does not hold for $d>2$.  

Another interesting finding was that in general one cannot use a higher dimensional multiport to simulate a lower dimensional one.  The significance of this is that one cannot use a single experimental setup to perform a variety of interference experiments.  We established this fact by showing that a three port interferometer (or indeed any interferometer with $d$ odd) could not be used to simulate a Mach-Zehnder interferometer.  Interestingly, a four port interferometer can be used to simulate a Mach-Zehnder interferometer, but it cannot simulate a three port interferometer.  It may be the case that a $d$ port interferometer can never simulate a $d-1$ port interferometer, while being able to simulate some other lower dimensional interferometers.  This question is, however, still unresolved. 

In all of the cases we have studied the input to the interferometer was a Fock state.  There are situations, such as continuous variable encoding, where we could have coherent states as inputs.  It has been shown \cite{bergou} that the transformation matrix for the input and output amplitudes of the coherent states is the related to the transformation matrix between the annihilation operators of the input and output modes.  One can thus directly apply the results found in sections \ref{secIII} and \ref{secIV} to determine what (unentangled\footnote{If the input states have no squeezing, then we can be sure that the output will be unentangled \cite{convar}.}) output states can be generated for a given coherent state input.  

\section*{Acknowledgement}
We acknowledge financial support from the Royal Society International Joint
Project grant 2006/R2–IJP and the Czech - Hungarian project grant MEB 041011 (CZ-11/2009); T.B., V.K. and I.J. also acknowledge financial support from the Doppler Institute and from grants MSM 6840770039 and MSMT LC06002 of the Czech Republic.

\end{document}